\documentclass[pra,superscriptaddress,showpacs,showkeys,twocolumn]{revtex4-1}
\usepackage{amsfonts,amsmath,amssymb,graphicx,epstopdf,verbatim,color,subfigure}
\usepackage[english]{babel}
\usepackage[normalem]{ulem}
\usepackage[toc,page]{appendix}

\newcommand{\vectr}[2]{ \left( \begin{array}{c}
#1 \\
#2 
\end{array} \right)}
\newcommand{\matrx}[4]{ \left( \begin{array}{cc}
#1 & #2 \\
#3 & #4 
\end{array} \right)}
\renewcommand{\Im}{\text{Im}}
\renewcommand{\vec}[1]{\mathbf{#1}}

\graphicspath{{./images/}}

\begin{document}

\title{Spontaneous Beliaev-Landau scattering out of equilibrium}
\author{Mathias Van Regemortel}
\affiliation{TQC, Universiteit Antwerpen, Universiteitsplein 1,
B-2610 Antwerpen, Belgium} 
%\affiliation{INO-CNR BEC Center and Dipartimento di Fisica, Universit\`a di Trento, via Sommarive 14, 38123 Povo, Italy}
\author{Wim Casteels}
\affiliation{TQC, Universiteit Antwerpen, Universiteitsplein 1,
B-2610 Antwerpen, Belgium} 
\author{Iacopo Carusotto}
\affiliation{INO-CNR BEC Center and Dipartimento di Fisica, Universit\`a di Trento, via Sommarive 14, 38123 Povo, Italy}
\author{Michiel Wouters}
\affiliation{TQC, Universiteit Antwerpen, Universiteitsplein 1,
B-2610 Antwerpen, Belgium}

\date{\today}
\begin{abstract}

 We investigate Beliaev-Landau scattering in a gas of interacting photons in a coherently driven array of nonlinear dissipative resonators, as described by the 1D driven-dissipative Bose-Hubbard model. Due to the absence of detailed balance in such an out-of-equilibrium setup, steady-state properties can be much more sensitive to the underlying microscopic dynamics. Because the popular truncated Wigner approximation dramatically fails in capturing this physics, we present an alternative approach, based on a systematic expansion beyond the Bogoliubov approximation, which includes the third-order correlation functions in the dynamics. As experimentally accessible signatures of Beliaev-Landau processes, we report a small but nonnegligible correction to the Bogoliubov prediction for the steady-state momentum distribution, in the form of a characteristic series of peaks and dips, as well as non-Gaussian features in the statistics of the cavity output field.

\end{abstract}

\maketitle

\section{Introduction}

The convenient assumption of detailed balance, valid for a quantum many-body system at thermal equilibrium, ensures that any microscopic process is balanced by its reverse process, thus making the specific underlying dynamics irrelevant for the equilibrium ensemble.
For systems far from equilibrium no such claims can be made. Although some similarities can exist \cite{sieberer_2016_keldysh, fossfeig_optical_bistability,Jose_Iacopo},  driving and dissipation in general prevent the system from approaching a complete thermal equilibrium \cite{carmichael_statistical}. As a result, the phase-space distribution of out-of-equilibrium systems is typically much more sensitive to the actual microscopic driving, dissipation and equilibration processes that drive the system towards the steady-state.

In this manuscript we study a weakly interacting driven-dissipative Bose-Hubbard model in the superfluid regime. Going beyond the widely applied Bogoliubov approximation, which assumes noninteracting quasiparticles, we show that scattering processes involving three quasiparticles, known as Beliaev-Landau scattering \cite{beliaev1958}, leave a small imprint on the steady-state momentum distribution and are responsible for non-Gaussian features in the photon statistics of the cavity output field.

Beliaev-Landau scattering processes were originally predicted in the many-body theory of quantum fluids~\cite{PinesNozieres} to be responsible for the finite lifetime of phonons in systems of either bosonic \cite{Pitaevskii_Stringari_BL,Giorgini_BL,Domokos_beliaev} or fermionic particles \cite{castin_landau} and were experimentally observed with ultracold atomic gases\cite{Foot_BL,Steinhauer_beliaev}. In these works, the damping was typically detected after a population of phonons was introduced in the system by an external perturbation. Here, on the contrary, we aim at capturing the spontaneous occurrence of Beliaev-Landau scattering processes by identifying their footprint on steady-state properties of the non-equilibrium photon gas, such as, most notably, the momentum distribution and the quantum statistics.

Semiconductor microcavities and superconducting circuits are among the most promising platforms for realizing scalable arrays of quantum-optical building blocks, suitable for large-scale quantum simulations (see \cite{schmidt2013,Iacopo_QFL,hartmann2016,angelakis2016} for recent reviews).  Motivated by recent experimental advances, much theoretical effort has been devoted to developing numerical techniques to simulate large-scale driven-dissipative quantum systems. Several reasons cooperate to make these systems computationally much more challenging than the corresponding equilibrium ones. First of all the total photon number is not conserved, resulting in a much larger effective Hilbert space, and secondly the steady state is a mixed state, thus requiring the evaluation of a full density matrix rather than a single wavefunction. This led to the development of new numerical tools such as, among others, variational approaches based on matrix product states in 1D \cite{verstraete_diss,orus_infinite,cui_variational,mascarenhas_matrix}, resummation techniques \cite{li_2016_resummation}, self-consistent projection operator theory \cite{degenfeld_2014_self}, extensions of the variational principle \cite{weimer_variational} and the corner-space renormalization method for 2D lattices \cite{finazzi_corner}.

In addition to the above-cited exact methods, approximated techniques based on the truncated Wigner approximation are also very popular tools to evaluate corrections beyond the Bogoliubov approximation in both conservative cold-atom \cite{castin_TWA,polkovnikov_phase_space} and lossy optical systems \cite{gardiner_quantum_noise, Iacopo_QFL}.
For the latter, this technique has been applied in various contexts, including the study of condensation and superfluid properties \cite{iacopo_spontaneous,wouters2009stochastic,gladilin2014spatial,he2015scaling}, dynamical phase transitions \cite{sieberer2013dynamical,dagvadorj_2015_nonequilibrium,fossfeig_optical_bistability}, and even genuine quantum effects such as the dynamical Casimir emission~\cite{koghee2014dynamical} and Hawking radiation~\cite{Hawking,Hawking2}. In contrast to these successes, we will show in this work that the truncated Wigner approximation, when naively adopted to study Beliaev-Landau scattering, may dramatically overestimate the corrections to Bogoliubov theory and even lead to unphysical results.

As an alternative approach we discuss how a truncated hierarchy of correlations can serve as a consistent expansion beyond the Bogoliubov approximation \cite{hartmann_2013_correlator,kastner_BBGKY,wim_hierarchy}. In particular, we will illustrate that the truncation of the hierarchy at the third-order correlation functions, i.e. one order beyond the Bogoliubov approximation, is sufficient to incorporate the corrections attributed to Beliaev-Landau processes, provided an adequate truncation scheme is employed.

The structure of our paper is as follows. In Sec. \ref{sec:model} we present the 1D driven-dissipative Bose-Hubbard model and in Sec. \ref{sec:beliaev} we illustrate how the non-equilibrium condition allows for on-shell Beliaev-Landau scattering in one-dimension. We next explain how the truncated Wigner approximation fails to describe these processes in Sec. \ref{sec:TWA}. In Sec. \ref{sec:HOC} we introduce the third-order correlation functions to incorporate Beliaev and Landau scattering and construct a hierarchy of correlation functions. In Sec. \ref{sec:exp} we discuss the imprint of Beliaev-Landau scattering on measurable quantities and discuss the expected signal for realistic parameters inspired from state-of-the-art semiconductor devices. Conclusions are finally drawn in Sec. \ref{sec:conclusion}. Appendix A reports additional 
numerical TWA data in the absence of open Beliaev-Landau channels. 
Appendices B and C summarize technical details on the hierarchy of 
correlations and on the different truncation schemes. Details on the 
calculation of the effect of disorder are given in Appendix D.

\section{The Model}
\label{sec:model}
We consider a 1D coupled array of $L$ nonlinear, single-mode photon cavities under a coherent drive with frequency $\omega_L$. The resonator frequencies $\omega_c$ are assumed to be uniform throughout the chain. After a unitary transformation to remove the time-dependence of the drive, we obtain the driven Bose-Hubbard Hamiltonian (we set $\hbar = 1$ throughout the article)
\begin{eqnarray}
\nonumber
\hat{H} &=& -J \sum_{\langle j,l\rangle} \left(\hat{a}^\dagger_j \hat{a}_{l} +\hat{a}^\dagger_l \hat{a}_{j}\right) - \delta \sum_{j=1}^L \hat{n}_j + \frac{U}{2} \sum_{j=1}^L \hat{n}_j(\hat{n}_j-1)\\ &&+ \sum_{j=1}^L\Omega_j ( \hat{a}_j + \hat{a}^\dagger_j).
\label{eq:Hamiltonian}
\end{eqnarray}
The operators $\hat{a}^\dagger_j(\hat{a}_j)$ create (annihilate) a particle at site $j$ of the chain and $\hat{n}_j = \hat{a}^\dagger_j\hat{a}_j$ is the local number operator. Photons in the chain can tunnel to their neighbouring sites with a hopping strength given by $J$. The notation $\langle j,l\rangle$ means that the summation runs over all neighbouring sites. The two-body interaction strength for photons confined inside the same cavity is given by $U$. The amplitude of the driving field at each site is $\Omega_j$, while its detuning from the onsite single-photon resonance is given by $\delta=\omega_L-\omega_c$. For simplicity we impose periodic boundary conditions, such that $\hat{a}_1=\hat{a}_{L+1} $.

The dissipative nature of the setup implies that injected photons have a finite lifetime inside the cavity array before they escape. In the Born-Markov approximation, the coupling of the system to its environment at zero temperature is described by the dissipator in the Lindblad form \cite{gardiner_quantum_noise,open_quantum_systems} 
\begin{equation}
\label{eq:dissipator}
\mathcal{D}[\hat{\rho}]=\frac{\gamma}{2}\sum_j(2\hat{a}_j\hat{\rho}\hat{a}^\dagger_j - \hat{\rho}\hat{n}_j - \hat{n}_j\hat{\rho}).
\end{equation}
The full dynamics of the density matrix $\hat{\rho}$ is then governed by a master equation, which includes both the unitary evolution under $\hat{H}$ and the photonic losses 
\begin{equation}
\label{eq:q_master}
\partial_t \hat{\rho} = -i[\hat{H},\hat{\rho}] + \mathcal{D}[\hat{\rho}]
\end{equation}

A mean-field description of the problem can be derived in terms of coherent fields $\psi_j = \langle \hat{a}_j \rangle$ by assuming that all normal-ordered operator products factorize~\cite{Iacopo_QFL}. This leads to the following equations of motion
\begin{equation}
\label{eq:MF}
i \dot{\psi}_j = -\Big(\delta + i\frac{\gamma}{2}\Big)\psi_j -J(\psi_{j+1} + \psi_{j-1}) + U|\psi_j|^2\psi_j + \Omega_j
\end{equation}

For this work we assume a uniform drive field $\Omega_j=\Omega$ at all sites in the chain. In the steady state we thus find one or two stable homogeneous density solutions in the mean-field description, depending on the amplitude and detuning of the pump. They are found as solutions of 
\begin{equation}
n_0((\delta - Un_0 + 2J)^2 + \gamma^2/4) = |\Omega|^2,
\end{equation}
where $n_0=|\psi_0|^2$ in terms of the spatially uniform mean-field steady state $\psi_j = \psi_0$. The parameter 
\begin{equation}
\label{eq:delta}
 \Delta= \delta - Un_0 + 2J
\end{equation} 
is the renormalized laser detuning from the interaction-blueshifted optical resonance. We restrict our analysis to the case $\Delta < 0$, such that the system is in the optical limiter regime or in the high-density branch of the hysteresis loop of a bistable regime \cite{Iacopo_QFL}.  As we will briefly review later, this restriction asserts a gapped spectrum of excitations \cite{Iacopo_probing}. In the numerical analysis that follows, the drive amplitude $\Omega$ is always implicitly determined by choosing a value for $n_0$.

\section{Bogoliubov theory and Beliaev-Landau processes}
\label{sec:beliaev}

\subsection{Bogoliubov dispersion and the non-equilibrium steady-state}

The case of a uniform drive field allows for a convenient parametrization of the full quantum field in terms of a homogeneous, coherent field $\psi_0$ and quantum fluctuations. Expanding the latter in their $\hat{\phi}_k$ spatial Fourier components, one can write
\begin{equation}
\label{eq:ansatz}
\hat{a}_{j} = \psi_0 + \frac{1}{\sqrt{L}} \sum_k e^{ikj} \hat{\phi}_k
\end{equation}
where the sum over $k$ is restricted to the interval $[-\pi,\pi]$ with a spacing equal to $2\pi/L$ in the case of a finite number $L$ cavities and periodic boundary conditions.

While the time evolution of the mean-field $\psi_0$ is governed by the classical equation (\ref{eq:MF}), the dynamics of the quantum fluctuations is governed by a quantum Langevin equation  \cite{gardiner_quantum_noise,Iacopo_QFL}
\begin{eqnarray}
\label{eq:q_fluct}
\nonumber
i\partial_t \hat{\phi}_k &=& (\epsilon_k + Un_0 - i\gamma/2)\hat{\phi}_k + U\psi_0^2 \hat{\phi}^\dagger_{-k} +  \hat{\xi}_k \\\nonumber
&&+ \frac{2U\psi_0}{\sqrt{L}} \sum_q \hat{\phi}^\dagger_q \hat{\phi}_{k+q} + \frac{U\psi_0^\ast}{\sqrt{L}} \sum_q \hat{\phi}_q \hat{\phi}_{k-q}\\
&&+ \frac{U}{L} \sum_{q,l} \hat{\phi}^\dagger_q \hat{\phi}_l \hat{\phi}_{k+q-l},
\end{eqnarray}
where we have set
\begin{equation}
\label{eq:epsilon}
\epsilon_k = -\delta + Un_0 - 2J\cos{k}.
\end{equation}
The Markovian losses are responsible for quantum noise with Gaussian statistics, represented by the operators $\hat{\xi}_k$, that assume the following zero-temperature statistics
\begin{eqnarray}
\big\langle\hat{\xi}_k(t)\hat{\xi}_{k'}(t')\big\rangle &=& \big\langle\hat{\xi}_k^\dagger(t)\hat{\xi}_{k'}(t')\big\rangle = 0,\\
 \big\langle \hat{\xi}_k(t)\hat{\xi}_{k'}^\dagger(t')\big\rangle &=& \gamma \delta_{k,k'} \delta(t-t').
\end{eqnarray}
As usual in Bogoliubov-like approaches, the interaction terms in (\ref{eq:q_fluct}) are ordered in increasing number of fluctuation operators. When the number of photons in the condensate $|\psi_0|^2$ is much larger than the number of fluctuations, one expects the effect of higher-order terms to be negligible~\cite{castin_dum}.

The first-order correction to the mean-field, summarized on the first line of (\ref{eq:q_fluct}), incorporates processes where two condensate particles collide and produce a pair of excitations with counter-propagating wavevectors $k$ and $-k$ and viceversa. In a Hamiltonian formalism, this corresponds to only retaining quadratic terms in the fluctuation operators $\hat{\phi}_k$. Restricting to these terms in (\ref{eq:q_fluct}) and dropping the ones on the second and third line, which contain terms with more than one fluctuation operator, results in a set of linear equations for the fluctuation fields $\hat{\phi}_k$. As done in~\cite{Iacopo_spec}, this set of linear stochastic equations is solved by means of a Bogoliubov transform to new operators
\begin{equation}
 \hat{\phi}_k = u_k\hat{\chi}_k + v_k\hat{\chi}^\dagger_{-k}\label{eq:bogo_transf}
\end{equation}
that diagonalize the equations of motion 
\begin{equation}
\label{eq:bog_dyn}
i(\partial_t + \gamma/2) \hat{\chi}_k = \omega_k \hat{\chi}_k + u_k \hat{\xi}_k - v_k \hat{\xi}^\dagger_{-k}.
\end{equation}
Here we have defined the quasiparticle energies $\omega_k$ along with the transformation functions $u_k$, $v_k$ as
\begin{eqnarray}
\label{eq:bog_transform}
\omega_k &=& \sqrt{\epsilon_k(\epsilon_k +2Un_0)},\\
u_k,v_k &=& \frac{\sqrt{\epsilon_k + 2Un_0} \pm \sqrt{\epsilon_k}}{2\sqrt{\omega_k}}.
\end{eqnarray}
In Fig. \ref{fig:bel_decay}(a) the Bogoliubov spectrum (\ref{eq:bog_transform}) is shown for different values of the renormalized detuning $\Delta$ (\ref{eq:delta}). In contrast to equilibrium systems, note that a spectral gap is generally present in the Bogoliubov dispersion and only closes for $\Delta \rightarrow 0^{(-)}$, i.e. when the drive is exactly on resonance with the interaction-shifted mode~\cite{Iacopo_QFL}.

 Due to the noise operators in (\ref{eq:bog_dyn}), one has a finite occupation of Bogoliubov modes with non-trivial anomalous correlations in the stationary regime,
\begin{equation}
\label{eq:bog_distr}
n^{(\chi)}_k =\langle \hat{\chi}_k^\dagger \hat{\chi}_k \rangle = v_k^2,\;\;\;c^{(\chi)}_k =\langle \hat{\chi}_k \hat{\chi}_k \rangle= \frac{u_k v_k \gamma}{\gamma + 2i\omega_k}
\end{equation}
It is important to note that the occupation of the Bogoliubov modes here, in contrast with an equilibrium system, is not at all set by a finite-temperature Boltzmann-Gibbs distribution but by the interplay of interactions, hopping, driving and dissipation~\cite{Iacopo_spec}. This different origin is apparent in the slow, power-law decay of the occupation of high momentum modes, much slower than the usual exponential $\exp{\big[-E(k)/k_BT\big]}$ of equilibrium systems.

Moving back to the original $\hat{\phi}_k$ operators, one can derive a closed system of linear differential equations for the quadratic correlation functions $n_k = \langle \hat{\phi}_k^\dagger \hat{\phi}_k \rangle$ and $c_k = \langle \hat{\phi}_k \hat{\phi}_{-k} \rangle$,
\begin{eqnarray}
\label{eq:2nd_bog}
\partial_t n_k &=& -\gamma n_k + 2\Im \big[U\psi_0^2 c^\ast_k\big]\\
i\partial_t c_k &=& (2\epsilon_k +2U|\psi_0|^2 - i\gamma)c_k + U\psi_0^2(2n_k + 1),
\end{eqnarray}
whose steady-state solution reads
\begin{equation}
\label{eq:momentum_distr}
n_k = \frac{1}{2}\frac{(Un_0)^2}{\omega_k^2 + \gamma^2/4},\;\;\; c_k = -\frac{U\psi_0^2}{2}\frac{\epsilon_k + Un_0 + i\gamma/2}{\omega_k^2 + \gamma^2/4}.
\end{equation}

\subsection{Corrections to the Bogoliubov approximation and Beliaev-Landau processes}
\begin{figure}
\centering
\includegraphics[width = \columnwidth]{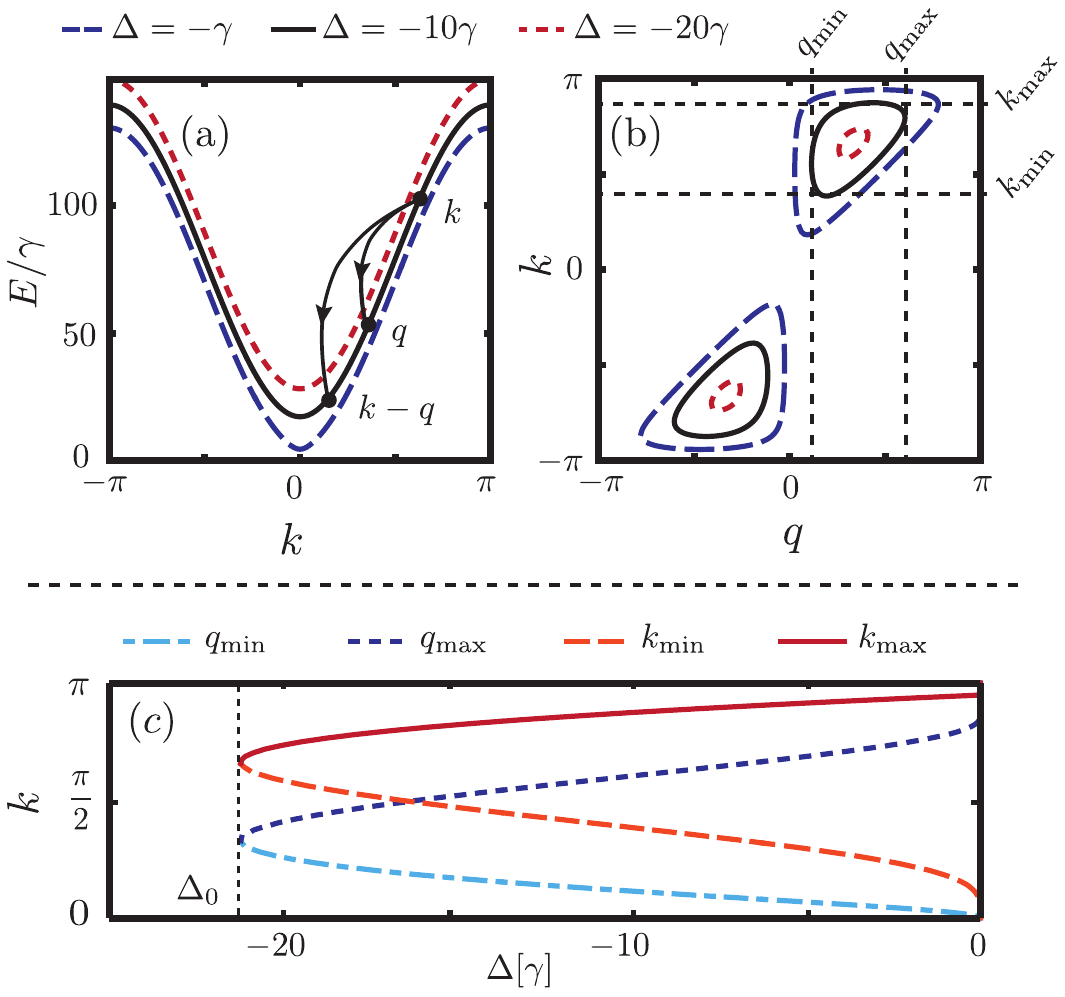}
\caption{ (a) The spectrum of excitations (\ref{eq:bog_transform}) for $(J, Un_0) = (30\gamma, 10\gamma)$ and for different values of the renormalized detuning $\Delta$ (\ref{eq:delta}). If $\Delta \rightarrow 0^{(-)}$, i.e. a drive exactly on the blue-shifted resonance, the gap vanishes. Beliaev decay is sketched as an excitation at momentum $k$ that decays to $q$ and $k-q$ with conservation of energy. (b) The contours of energy conservation from (\ref{eq:E_cons}) for the same parameters as panel (a). The extremal momenta are found from the contours and are indicated for $\Delta = -10\gamma$ (full black line), the case we have considered for the rest of the analysis. (c) The shift of the extremal momenta as a function of $\Delta$ for the positive contour. At $\Delta<\Delta_0<0$ the spectral gap becomes too large and no resonant scattering channels exist.  }
\label{fig:bel_decay}
\end{figure}

The next-order correction to the Bogoliubov approximation is given by Hamiltonian interaction terms that comprise three fluctuation operators and only one condensate mode, which go under the name of Beliaev and Landau scatterings. Beliaev scattering is the collision of a fluctuation with momentum $\vec{k}$ with a condensate particle into a pair excitations with momenta $\vec{q}$ and $\vec{k}-\vec{q}$, such that total momentum is conserved (see Fig. \ref{fig:bel_decay} (a)). Landau scattering is the opposite process: two fluctuations with momenta $\vec{k-q}$ and $\vec{q}$ scatter into a condensate mode and an excitation with momentum $\vec{k}$.

In a closed system, all scattering processes must occur on-shell, i.e. they conserve both energy and momentum. For the Beliaev-Landau processes, this implies the following relation:
\begin{equation}
\label{eq:E_cons}
\omega_\vec{k}  = \omega_\vec{q} +  \omega_{\vec{k} - \vec{q}},
\end{equation}
with $\omega_k$ the quasiparticle oscillation frequency.
Because of the absence of a spectral gap and the convexity of the Bogoliubov dispersion of conservative continuum systems, criterium (\ref{eq:E_cons}) in continuum models can only be satisfied in two or more spatial dimensions, while in 1D sytems phonons can only decay through higher-order scattering processes~\cite{ristivojevic_2016_decay}. However, there exist specifically engineered 1D optical lattices with a nonconvex (but gapless) spectrum, such that energy and momentum conservation can be simultaneously satisfied \cite{optical_lattice_griffin,optical_lattice_tetsuro}.

The situation is different in driven-dissipative systems, where the Bogoliubov spectrum is typically gapped when the drive is below resonance. In a continuous 1D setup, the presence of a finite spectral gap in combination with a convex excitation spectrum always allows for third-order scattering process that satisfy (\ref{eq:E_cons}).  When the spectrum is not convex, as is the case in a lattice model, the situation is somehow more complicated. The subtle interplay between the spectral gap and the degree of nonconvexity determines whether resonant third-order scattering channels are present.

The allowed wavevectors $k$ and $q$ that exactly satisfy the energy and momentum conservation condition (\ref{eq:E_cons}) in one-dimension are indicated in Fig. \ref{fig:bel_decay}b). Importantly, maximal and minimal values can be deduced for $k$ and $q$ from the contours, which set limits on allowed in and out states for Beliaev-Landau scattering. Only excitations with a wavevector $k$ for which $k_\text{min} < k < k_\text{max}$ can scatter resonantly to excitations with wavevectors $q$ and $k-q$ through Beliaev decay. Likewise, only excitations with a wavevector $q$ for which $q_\text{min} < q < q_\text{max}$ can combine with an excitation at $k - q$ to form one at $k$ through Landau scattering. However, the driven-dissipative nature of our setup allows energy not to be strictly conserved, so that scattering processes are possible within a finite linewidth $\gamma$ around the energy-conservation point.

In Fig. \ref{fig:bel_decay}c) we show how the extremal input and output momenta shift as a function of the interaction-renormalized detuning, defined in $\Delta$ (\ref{eq:delta}), a parameter that can be tuned in experiment by changing the laser frequency $\omega_L$. When the drive is too far below resonance, i.e. when $\Delta < \Delta_0$, with $\Delta_0<0$ a critical value that can be derived from the dispersion relation, the spectral gap is too large as compared to the bandwidth and no resonant Beliaev-Landau scattering channels exist. In the limit of $\Delta \rightarrow 0$, for which the dispersion relation is linear, we find the contour of an equilibrium condensate from Ref. \cite{optical_lattice_griffin}. We anticipate at this point that the experimental possibility of shifting the limiting scattering momenta in a well-controlled manner provides a genuine signature of Beliaev-Landau scattering. 

Before continuing with our analysis, we would like to draw attention to an important consideration. Given a closed quantum system (e.g. a gas of ultracold atoms or a superfluid liquid Helium sample), the presence of detailed balance will unavoidably restrict the effect of Beliaev-Landau scattering to driving the system back into its thermal state once it is kicked out of equilibrium. Therefore most works on this physics are related to phonon-decay experiments, where one studies how externally injected phonons are damped through scattering with the condensate (Beliaev) or with the thermal cloud (Landau) \cite{Pitaevskii_Stringari_BL,Giorgini_BL} or how a thermal equilibrium is reached again after a sudden global quench \cite{Menegoz_BL}.

Consequently, it is exactly the \emph{absence} of detailed balance in a driven-dissipative context which motivates us to study the effects of spontaneous Beliaev-Landau processes in the steady-state regime of the cavity array. In this section we have illustrated that two crucial conditions for these scatterings to be possibly relevant are indeed satisfied: 1) there is finite occupation of Bogoliubov modes over the entire Brillouin zone, as given in (\ref{eq:bog_distr}), and 2) there are regions in phase space for which energy and momentum are conserved according to (\ref{eq:E_cons}), which allows the Bogoliubov modes to scatter and redistribute (quasi)resonantly.

\section{The Truncated Wigner Method}
\label{sec:TWA}

A possible approach to compute corrections beyond Bogoliubov and quantify the observable signatures of the Beliaev-Landau processes is the so-called Truncated Wigner approximation (TWA) \cite{gardiner_quantum_noise,vogel_twa,castin_TWA,polkovnikov_phase_space}. This approach is based on a one-to-one mapping of the quantum master equation for the density matrix (\ref{eq:q_master}) onto a partial differential equation for the corresponding Wigner distribution. 
The resulting Fokker-Planck equation can be efficiently simulated if the terms with a third-order derivative are neglected. Since these terms are 
proportional to the single-particle interaction constant $U$, one 
expects this approximation to be accurate for sufficiently weak values 
of $U$~\cite{Iacopo_QFL}. This leads to a stochastic differential equation for a classical field $\varphi_j(t)$ 
\begin{equation}
\label{eq:TWA}
\begin{split}
i d\varphi_j(t) = \Big[-\Big(\delta + i\frac{\gamma}{2}\Big)\varphi_j(t) -J\big(\varphi_{j+1}(t) + \varphi_{j-1}(t)\big)\\
 + U\big(|\varphi_j(t)|^2-1\big)\varphi_j(t) + \Omega_j(t) \Big]dt + \sqrt{ \frac{\gamma}{2}} dW_j(t),
\end{split}
\end{equation}
where the stochastic Wiener increment $dW_j(t)$ is white Gaussian noise with variance $\big\langle dW^\ast_j(t) dW_{j'}(t) \big\rangle = \delta_{j,j'} dt$ and a random phase. Average values of the field $\varphi_j$ correspond to expectation values of symmetrically ordered products of quantum operators. In particular, for the number operator we find
\begin{equation}
\label{eq:TWA_dens}
\big\langle\varphi_j^\ast \varphi_j\big\rangle_W = \frac{1}{2}\Big( \langle \hat{a}_j^\dagger \hat{a}_j \rangle +  \langle \hat{a}_j \hat{a}_j^\dagger \rangle \Big) = \langle \hat{n}_j \rangle + \frac{1}{2}.
\end{equation}

As a consequence, the quantum vacuum is represented by a finite occupation of $1/2$ for the classical field $\varphi_j$. As long as $n_j \gg 1/2$, one does not expect this to cause problems but, when performing a TWA simulation to estimate the effects of Beliaev-Landau scattering, one finds surprisingly large corrections to the occupation numbers of quantum fluctuations, as can be seen in Fig. \ref{fig:2nd}a).  Even worse is that the occupation numbers may become negative at certain values of the momentum, even though the used parameters are well inside the supposed region of validity of TWA. In Fig. \ref{fig:2nd} we fixed $Un_0 = 10\gamma$ and show the results for two interaction constants $U=0.02\gamma$ and $U=0.1\gamma$, such that the mean-field predictions for the number of particles per site are $n_0 = 500$ and $n_0 = 100$, respectively. The results were obtained by averaging out over a total number of about $10^6$ samples, which were collected by integrating (\ref{eq:TWA}) in time with small enough time step $\Delta t$, and then taking a statistically independent sample each $\tau_s = 5\gamma^{-1}$.

As expected, the magnitude of the correction to the Bogoliubov theory is proportional to the single-photon interaction constant $U$ (or, equivalently, to the inverse of $n_0$ at a given mean-field energy $Un_0$). For both values of $U$, the negative occupation of some high-$k$ modes is a clearly unphysical prediction of the TWA. 

To better understand the physical origin of this breakdown, one needs to take a closer look at the nature of the underlying physical processes. Through Beliaev scattering, a quasiparticle at a high momentum $k$ decays into two quasiparticles with smaller momenta $q$ and $k-q$. Since the occupation decreases for larger $k$-modes, one expects the importance of this effect to be suppressed at higher momenta. However, within the TWA the quantum field is represented as a classical field for which the occupation of high-$k$ modes does not decay to $0$ but to $1/2$, which represents the quantum vacuum fluctuations (see Eq. (\ref{eq:TWA_dens})). This finite occupation of all modes, even the highest-$k$ ones, results in the possibility of a nonphysical decay of the quantum vacuum through spontaneous Beliaev processes. The final states of these collisions are quasiparticles with smaller momenta, which explains the massive pileup in the momentum distribution around $q_\text{min}$, at the cost of a strong negative dip around $k_\text{max}$. Of course, the TWA-simulated momentum distribution recovers relatively well to the Bogoliubov result for all $k$ values outside of the region $[q_\text{min},k_\text{max}]$ for which there are no resonant Beliaev-Landau scatterings possible. In Appendix \ref{app:TWA} we include a simulation of a model without energy-conserving Beliaev-Landau channels and we conclude that in this case the occupation of all modes is positive and much better convergence to the Bogoliubov result is achieved.  While this inaccuracy of the TWA is not expected to affect the predictions for dynamical Casimir and Hawking emission~\cite{koghee2014dynamical,Hawking,Hawking2} that are at the level of Bogoliubov theory, special care will be needed in the more advanced study of back-reaction effects in analog models of gravity \cite{balbinot_br}.

As far as we we know, there is no simple solution to this intrinsic problem of the TWA. Note that related problems with the TWA are known also in the conservative case of ultracold atomic Bose gases \cite{castin_TWA}. The equipartition theorem for the classical fields states in fact that the momentum distribution should eventually relax to a thermalized one satisfying $n_k^\text{class} \sim k_BT/\epsilon_k$. Apart from the fact that this Rayleigh-Jeans-like law does not match the expected Bose statistics, the TWA can also lead to negative values for the physical occupation of high-$k$ modes after subtraction of $1/2$ vacuum noise (\ref{eq:TWA_dens}). Therefore one can reliably use the TWA to compute time evolutions only up to a limited time, such that no thermalization sets in for the high momentum modes. Also on the calculation of phonon damping rates this problem has a direct impact, as the unphysical late-time 
thermalization of the classical field dramatically affects the Landau 
processes \cite{castin_TWA}.

\section{Hierarchy of correlation functions}
\label{sec:HOC}

Given the dramatic failure of the TWA classical field approach 
discussed in the previous section, we need to develop a more 
sophisticated method, apt to capture the quantum nature of the field operators more accurately. The idea is to go back to the quantum equation of motion (\ref{eq:q_fluct}) and to expand up to higher orders in the quantum fluctuations. 

\subsection{The method}

\begin{figure}
\centering
\includegraphics[width = \columnwidth]{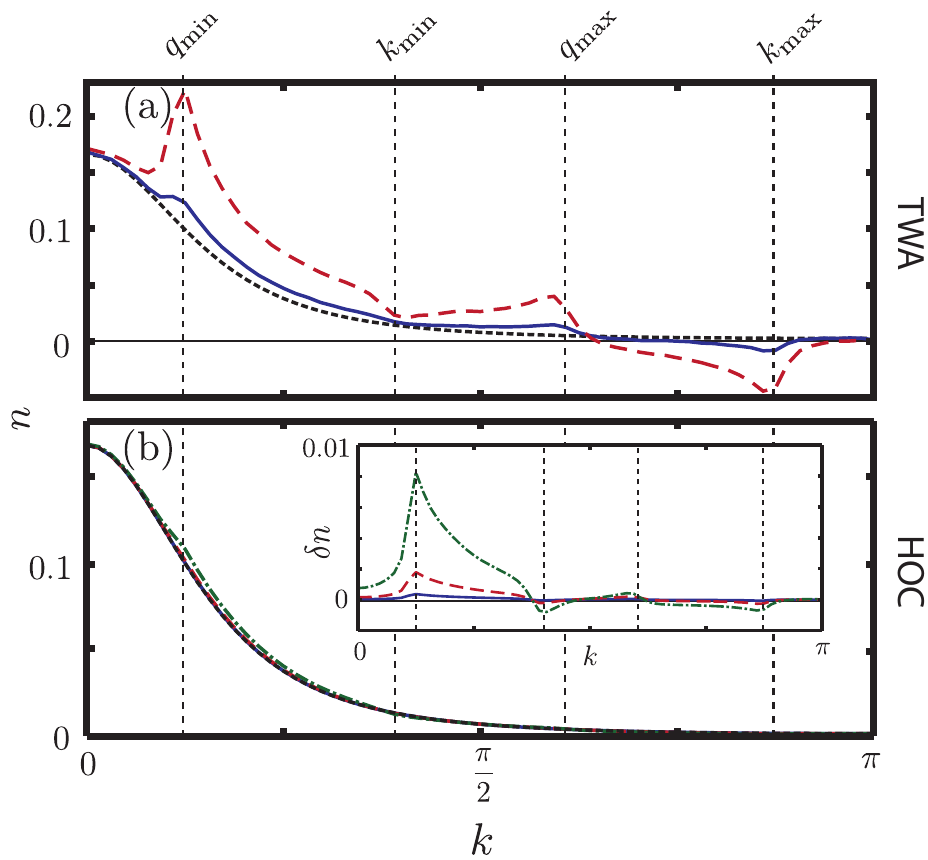} 
\caption{The momentum distribution of photons (in units of number of photons per mode) as obtained (a) from the truncated Wigner approximation--TWA and (b) by truncating the hierarchy of correlation functions--HOC at the third-order as discussed in the text. The parameters are $(J,\Delta, Un_0) = (30\gamma,-10\gamma, 10\gamma)$, $ L=128$ and three different interaction strengths $U=0.02\gamma$ (blue full lines), $U=0.1\gamma$ (red dashed lines) and $U=0.5\gamma$ (green dash-dotted lines), such that the average number of photons per cavity is $500$,$100$ and $20$ respectively. The latter case is not shown for the TWA computation, because it is outside of its regime of validity anyway.  The Bogoliubov result (\ref{eq:2nd_bog}) is also indicated (black dotted line). For clarity, we show the difference of the HOC with Bogoliubov $\delta n = n - n_\text{bog}$ in the inset of (b). The redistribution of particles is in both cases the strongest around the extremal values $q_{min}$, $k_{min}$, $q_{max}$ and $k_{max}$ (vertical dotted lines) of the contour from Fig. \ref{fig:bel_decay}b). The inaccurate TWA result tremendously overestimates the corrections stemming from Beliaev-Landau decay, with a negative value for certain $k$ modes, while the HOC result predicts only a small deviation from the Bogoliubov distribution. This is a direct consequence of the unphysical decay of the quantum vacuum in the Wigner representation.}
\label{fig:2nd}
\end{figure}
While a linearized form of eq. (\ref{eq:q_fluct}) was sufficient to reproduce the quadratic correlation functions (\ref{eq:bog_distr}), which describe the effect of a nonzero quasiparticle occupation, one can expect that the third-order correlation functions are needed to correctly describe interactions between quasiparticles. In particular, the matrix $M_{k,q}^{(\chi)} = \langle \hat{\chi}_{k-q}^\dagger \hat{\chi}_{q}^\dagger \hat{\chi}_{k} \rangle$ can be used to represent the scattering of a quasiparticle with momentum $k$ into two quasiparticles with momenta $q$ and $k-q$ and vice versa, i.e. Beliaev and Landau scattering. 
To facilitate our discussion, from now on we go back from the Bogoliubov basis to the original basis of $\hat{\phi}_k$ operators. This requires including two distinct third-order correlators in the dynamics, namely
\begin{equation}
\label{eq:3rd_corrs}
M_{k,q} = \langle \hat{\phi}_{k-q}^\dagger \hat{\phi}_{q}^\dagger \hat{\phi}_{k} \rangle,\;\;\; R_{k,q} = \langle \hat{\phi}_{-k-q} \hat{\phi}_{q} \hat{\phi}_{k} \rangle.
\end{equation}
Making use of equation (\ref{eq:q_fluct}) for the time evolution of quantum fluctuations, one readily derives differential equations for the correlation functions up to third order (see Appendix \ref{app:HOC}). 

If the lowest order $\psi_0$ in Eq. (\ref{eq:ansatz}) is kept fixed to the mean-field value, the inclusion of the third-order correlator leads to a finite value for the first-order correlator $\phi_0=\langle \hat{\phi}_0 \rangle$ as well. Another convenient way of choosing the ansatz (\ref{eq:ansatz}) is to set $\phi_0 = 0$ by definition, thus capturing the variation of the condensate wavefunction directly in $\psi_0$. This goes at the cost of adding back-reaction terms to the generalized Gross-Pitaevskii equation. At the level of approximation considered in this section, both approaches are equivalent, but setting $\phi_0 = 0$ asserts that we are dealing with connected second- and third-order correlation functions, which is a better controlled truncation \cite{kohler_microscopic}. We refer the interested readers to Appendix \ref{app:HOC} for more details on the method and to Appendix \ref{app:comp} for a comparison of different truncation schemes.

Correlation functions of order four, which enter into the equations of motion of the second and third-order correlators, are factorized into different possible products of second-order correlation functions. With this procedure we explicitly neglect the connected part of the fourth-order correlation function, but we keep its main contribution coming from separable correlations.
The fifth-order correlator, entering in the equation of motion for the third-order correlation functions, can be instead safely neglected. Already in factorized  form it would reduce to various products of second and third order, which constitute negligible corrections to dominant terms in the equations of motion. See Appendix \ref{app:HOC} for more details on the implications of these approximations.

Within this framework, the Gross-Pitaevskii equation for the homogeneous condensate background $\psi_0$, extended with the back-reaction terms reads
\begin{eqnarray}
i \partial_t \psi_0 &=& \Big(-\Delta - i\frac{\gamma}{2}\Big)\psi_0   + \Omega\\\nonumber
&& + \frac{2U\psi_0}{L}\sum_k n_k + \frac{U\psi_0^\ast}{L}\sum_k c_k + \frac{U}{\sqrt{L^3}} \sum_{k,q} M^\ast_{k,q}.
\label{eq:HE_MF}
\end{eqnarray}
The second-order correlation functions (\ref{eq:2nd_bog}) are now coupled to the third-order correlation functions (\ref{eq:3rd_corrs})
\begin{eqnarray}
\nonumber
\label{eq:HE_n}
i\partial_t n_k &=& -i\gamma n_k +  2i\Im\bigg[ U\Big(\psi_0^2 + \frac{1}{L}\sum_q c_q \Big) c^\ast_k \\
&& + \frac{2U\psi_0}{\sqrt{L}} \sum_q  M_{q,k} + \frac{U\psi_0^\ast}{\sqrt{L}} \sum_q M^\ast_{k,q} \bigg],\\
\nonumber
\label{eq:HE_c}
i\partial_t c_k &=& \bigg(2\epsilon_k+2U\Big( |\psi_0|^2 +\frac{1}{L} \sum_q n_q\Big) -i\gamma\bigg) c_k \\\nonumber
&&+ U\Big( \psi_0^2 + \frac{1}{L} \sum_q c_q \Big)(2 n_k + 1)\\\nonumber
&&+\frac{2U\psi_0}{\sqrt{L}}\sum_q\left(M_{q,-k}^\ast + M_{q,k}^\ast\right) \\
&&+ \frac{U\psi_0^\ast}{\sqrt{L}}\sum_q \left( R_{-k,q } + R_{k,q} \right).
\end{eqnarray}
Note that the factorized contribution of the fourth-order correlator enters here in the equations of motion as a small correction to the couplings $\psi_0^2$ and $|\psi_0|^2$.  At equilibrium these corrections are well-studied in the Hartree-Fock-Bogoliubov method \cite{andersen}.

Finally we also find the equations of motion for the third-order correlation functions, in which the fifth-order back-reaction is neglected
\begin{eqnarray}
\nonumber
i\partial_t M_{k,q} &=& \left( \epsilon_k - \epsilon_q  - \epsilon_{k-q} - U|\psi_0|^2-\frac{3i}{2}\gamma \right) M_{k,q} \\\nonumber
&&- U\big(\psi_0^\ast\big)^2\left(M^\ast_{q,k} + M^\ast_{k-q,k} \right) \\
&&+ U\psi_0^2\; R^\ast_{-k,q} +  F^{(M)}_{k,q} , \label{eq:HE_M}\\
\nonumber
i\partial_t R_{k,q} &=& \left( \epsilon_k +\epsilon_q + \epsilon_{k+q} + 3U|\psi_0|^2 -\frac{3i}{2}\gamma\right) R_{k,q} \\\nonumber
&&+ 
U\psi_0^2\left(M^\ast_{-k,q}+M^\ast_{-q,k} + M^\ast_{k+q,k} \right)\\
&& + F^{(R)}_{k,q}.
\label{eq:HE_R}
\end{eqnarray}
Here $F^{(M,R)}_{k,q}$ captures the back-reaction of the various separable contributions from the fourth-order correlation functions
\begin{eqnarray}
\nonumber
F^{(M)}_{k,q}&=& \frac{2U\psi_0}{\sqrt{L}} \Big(c^\ast_{k-q}n_q + n_{k-q}c^\ast_q - n_k(c^\ast_q+c^\ast_{k-q})\Big)\\\nonumber
&&+ \frac{2U\psi_0^\ast}{\sqrt{L}}\Big(n_{k-q}n_q-n_k(1+n_q +n_{k-q})\\
&& \;\;\;- c_k(c^\ast_q+c^\ast_{k-q})\Big)  \label{eq:F_M}\\\nonumber
F^{(R)}_{k,q}&=& \frac{2U\psi_0}{\sqrt{L}}\Big(c_k + c_q + c_{k+q} + n_{k+q}c_q + c_{k+q}n_q\\\nonumber
&& \;\;\; + n_k c_q + n_k c_{k+q} + c_k n_q + c_k n_{k+q} \Big)\\
&&+\frac{2U\psi_0^\ast}{\sqrt{L}}\Big(c_k c_q + c_k c_{k+q} + c_q c_{k+q} \Big)
\label{eq:F_R}
\end{eqnarray}
 In principle equations (\ref{eq:HE_MF}-\ref{eq:HE_R}) provide a solution to the full time-dependent problem when appropriate initial conditions are inserted. The focus of the present work is, however, on the steady-state solution. To obtain this, we in practice initialize the system with the mean-field condensate amplitude $\psi_0$ and the Bogoliubov solution (\ref{eq:momentum_distr}) for $n_k$ and $c_k$, and we initially set $M_{k,q}$ and $R_{k,q}$ to zero. We then let the system evolve until it spontaneously reaches its steady-state.

To follow the time-evolution, we have implemented a Runge-Kutta-based routine with adaptive timestep to integrate equations (\ref{eq:HE_MF}-\ref{eq:HE_R}) in time. By plotting a quantity such as $ \delta(t) = 1/(L\Delta t)\sum_k{ \big| n_k^{t + \Delta t} - n_k^t\big| / n_k^t} $ as a function of $t$ for fixed $\Delta t$, we can monitor the convergence. As a criterion we set a fixed $\epsilon$ and stop the evolution once $\delta(t) < \epsilon$. Typically $\delta(t) \sim \exp( -\kappa t)$ and therefore convergence is rapidly achieved. For a system with 128 cavities we need about 2 minutes of CPU time on a standard computer, without any optimization, to get an accuracy $\delta < 10^{-6}$.  

\subsection{Results}

In Fig. \ref{fig:2nd} we present a comparison between the (inaccurate and unreliable) TWA result and the one obtained with the present hierarchy-of-correlations (HOC) approach. First of all, the corrections from the Bogoliubov theory are again as expected proportional to the single-photon interaction constant $U$ for a given mean-field energy $Un_0$.  Importantly, the predictions of the HOC do not suffer from unphysical negative occupation numbers and quantitatively the corrections turn out to be much smaller than the ones found in TWA. From a qualitative point of view, we see that they are similar in shape to the TWA ones, but far less pronounced. We therefore conclude that within the TWA, the physical scattering processes between quasiparticles are overwhelmed by the unphysical Beliaev-like decay of the $1/2$ vacuum noise, which is indeed significantly larger than the actual occupation number of the excitations.

\begin{figure}
\centering
\includegraphics[width = \columnwidth]{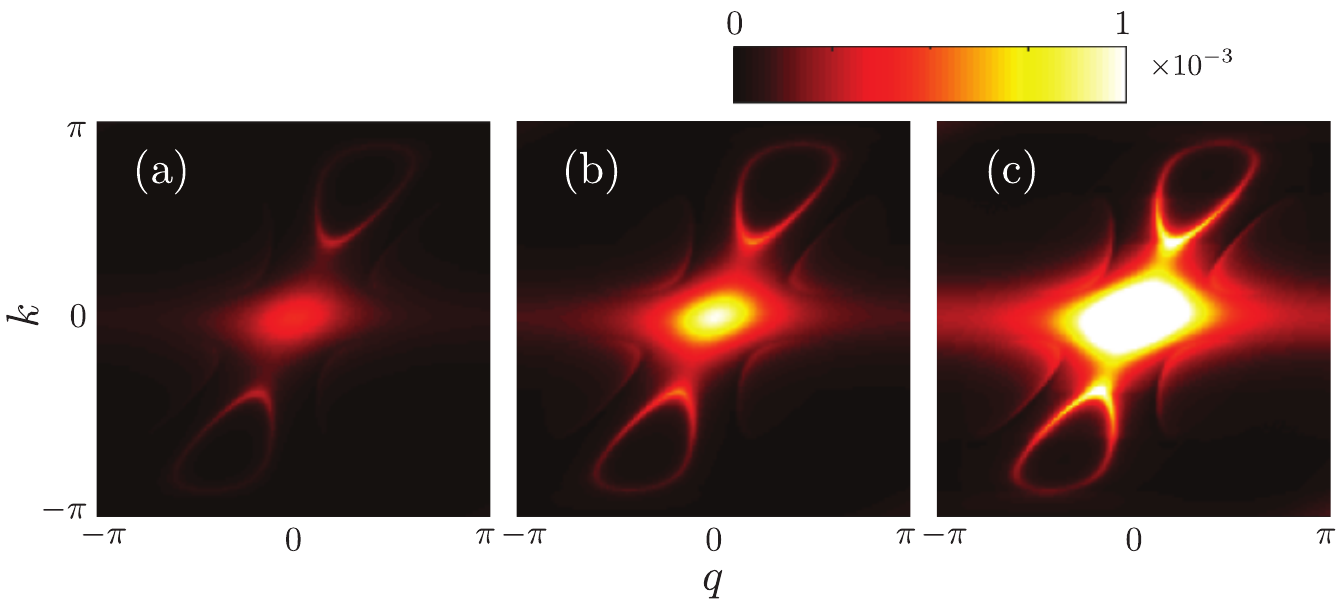}
\caption{The absolute value of the third-order correlation matrix $M_{k,q}$ for $U=0.02\gamma$ (a), $U = 0.1\gamma$ (b) and $U = 0.5\gamma$ (c) for the same parameters as Fig. \ref{fig:2nd}. The curve representing energy and momentum conservation shown as a full black line in Fig. \ref{fig:bel_decay}b) corresponds here to a line of enhanced scattering. }
\label{fig:3rd}
\end{figure}

Furthermore, we deduce from Fig. \ref{fig:2nd}b) that the overall redistribution of particles is from high $k$ to small $k$, which means that the Beliaev decay of high-momentum quasiparticles is dominant. Around the two extremal momenta of the input states, $q_\text{min}$ and $q_\text{max}$, this manifests itself as a peak in the momentum distribution, while there is a dip at the extremal momenta $k_\text{min}$ and $k_\text{max}$ of the output states. This is a consequence of the relatively large density of states for possible output (input) states for scatterings with input (output) momenta around $k_\text{min}$ or $k_\text{max}$ ($q_\text{min}$ or $q_\text{max}$), as one can deduce in Fig.\ref{fig:bel_decay}b from the slow bending of the contours at these extremal values. As energy does not need to be exactly conserved in an open system, scattering is also possible slightly outside the interval $[q_\text{min},k_\text{max}]$, with a width set by the linewidth $\gamma$. This characteristic series of peaks and dips in the steady-state momentum distribution appears to be a promising experimental signature of Beliaev-Landau scattering processes in a novel context of non-equilibrium quantum fluids. 
On the other hand, in the limit of small and large momenta ($k \rightarrow 0$ and $k \rightarrow \pi$) Beliaev-Landau processes are not allowed, so the Bogoliubov result is accurately recovered. 

When trying to gain insight into the nature of out-of-equilibrium Beliaev-Landau scattering, it is worthwhile to take a closer look at the scattering matrix $M_{k,q}$, shown in Fig. \ref{fig:3rd}. In addition to a central peak as a consequence of nonresonant decay, the contour representing energy and momentum conservation (see Fig. \ref{fig:bel_decay}b)) is clearly manifested as a band of enhanced magnitude of $M_{k,q}$.

To clarify this, we take a step back and go again to the basis of Bogoliubov operators $\hat{\chi}_k$. By pursuing transformation  (\ref{eq:bogo_transf}) consistently, we find a closed set of equations equivalent to (\ref{eq:HE_MF}-\ref{eq:HE_R}), but in terms of the $\hat{\chi}_k$. Although the full evaluation is much more cumbersome, as a consequence of the appearance of various products of the $u_k$ and $v_k$ transformation functions, one easily sees that the third-order correlation function must be of the form
\begin{equation}
\label{eq:M_bog}
\langle \hat{\chi}_{k-q}^\dagger \hat{\chi}_{q}^\dagger \hat{\chi}_{k} \rangle =\\ \frac{2U}{\sqrt{L}}\dfrac{\psi_0 A_{k,q} + \psi_0^\ast B_{k,q}}{\omega_k - \omega_q - \omega_{k-q} - \frac{3i}{2}\gamma},
\end{equation}
where the $A_{k,q}$ and $B_{k,q}$ are coefficients of order one which result from the combination of the Bogoliubov $u_k,v_k$ factors corresponding to the different terms originating from the factorization of the fourth-order correlation functions in the Bogoliubov basis.

From the denominator of expression (\ref{eq:M_bog}) one readily concludes that Beliaev-Landau scatterings are concentrated around the energy-conserving contours from (\ref{eq:E_cons}). As the hierarchy of correlations (\ref{eq:HE_MF}-\ref{eq:HE_R}) is built in the basis of the $\hat{\phi}_k$ operators, scatterings to negative energy states are also possible through the Bogoliubov transformation (\ref{eq:bogo_transf}). These contours can be obtained by setting $\omega_k \rightarrow -\omega_{-k}$ and/or $\omega_q \rightarrow -\omega_{-q}$ in (\ref{eq:E_cons}) and are visible as less pronounced bands of enhanced matrix elements in Fig. \ref{fig:3rd}. 

 While the Bogoliubov approximation is a consistent expansion beyond mean-field that captures corrections which scale as $\sim U\psi_0^2$, we now conclude from (\ref{eq:M_bog}) that terms scaling as $\sim U\psi_0/\sqrt{L}$, the next order in the expansion, are included with the present method. In particular, we have shown that the redistribution of occupation numbers is caused by quasiresonant Beliaev-Landau scattering.  In our framework, this effect is accounted for in (\ref{eq:HE_n}) and (\ref{eq:HE_c}) by the back-reaction of the third-order correlation matrices, which capture the scatterings, on the momentum distribution and anomalous correlation.

 \section{Experimental signatures of Beliaev-Landau scatterings}
 \label{sec:exp}
\begin{figure}
\centering
\includegraphics[width = .8\columnwidth]{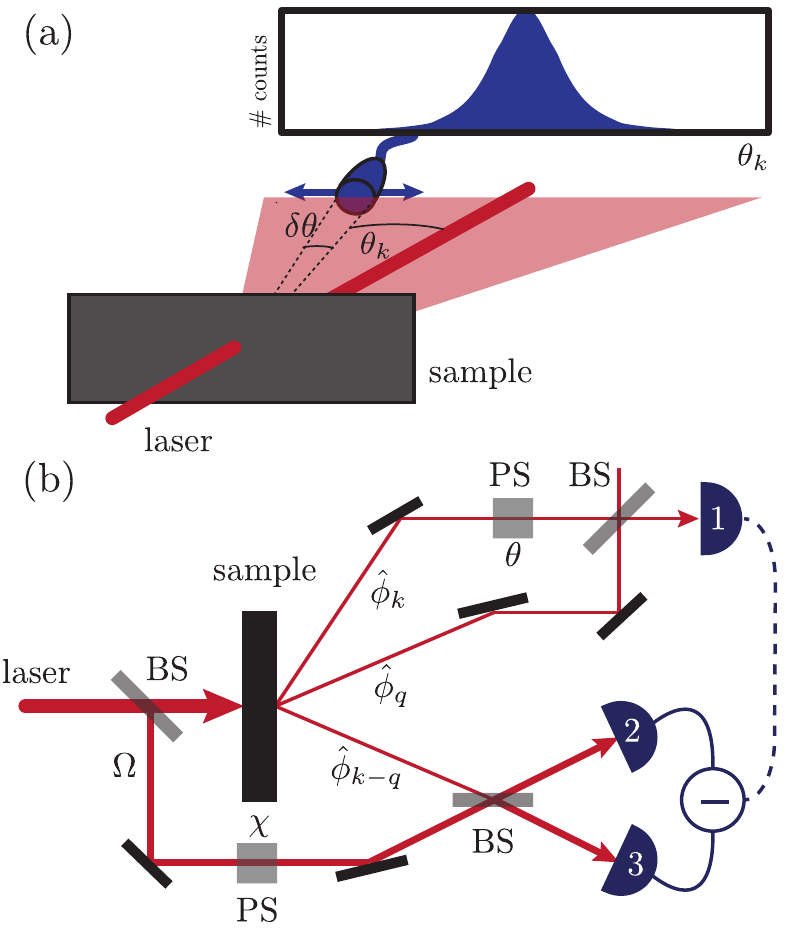}
\caption{(a) A sketch of the standard setup for an angle-resolved measurement. While the cavity array is pumped in $k=0$ mode, corresponding to a perpendicularly incident laser beam, quantum fluctuations are expected to leak out at a nonzero angle $\sin{\theta_k} = ck /(\omega_L\Delta x)$. In the setup we discuss in the text, all photons are expected to leak out within an angle of about $23^\circ$.  Measuring the intensity at an angle $\theta_k$ with width $\delta \theta$ then allows for the measurement of $n_k$, for which the theoretical prediction is given in Fig. \ref{fig:2nd}b). (b) A possible setup to detect the non-Gaussianities in the cavity output field through the third-order correlator (\ref{eq:3rd_corrs}). Simultaneous clicks between detector 1 and the difference signal of detectors 2 and 3 allows for the measurement of the quantity from expression (\ref{eq:detection}). 'PS' stands for phase shifter and 'BS' for a (50:50) beam splitter. }
\label{fig:setup}
\end{figure}

To make the theoretical analysis discussed in the previous sections more concrete, we dedicate this section to providing some clear experimental indications on how the small Beliaev-Landau signal can be extracted in a realistic experimental context. We consider a one-dimensional cavity array consisting of a chain of coupled semiconductor microcavities, such as presented in Ref. \cite{bloch_simulator}. The cavities, $L$ in total, are positioned at a distance  $\Delta x=1$ $\mu$m from each other and are irradiated by a laser with frequency $\hbar \omega_L = 1.6$ eV. Photons in the cavities have an average lifetime of $20$ ps, corresponding to a linewidth of $\hbar\gamma = 33$ $\mu$eV and display a single-photon nonlinearity of $U = 3.3$ $\mu$eV, such that $U \approx 0.1 \gamma$. Furthermore we set $J=30\hbar\gamma \approx 1m$eV, $\Delta = -10\hbar\gamma\approx -330\mu$eV and assume an average number of photons per cavity of $n_0=100$, such that $Un_0 \approx 10\gamma \approx 330\mu$eV. This is a case that we have already discussed in the theoretical analysis presented in Sec. \ref{sec:HOC} (see Fig. \ref{fig:2nd}b red dashed line and Fig. \ref{fig:3rd}b)).

As we have illustrated previously, the most straightforward approach to observe a signature of Beliaev-Landau scattering in the cavity array is to measure the momentum distribution of the quantum fluctuations and observe the characteristic series of peaks and dips. With the proposed parameters, we predict that a deviation of about $2\%$ from the Bogoliubov result (\ref{eq:momentum_distr}) can be observed around the minimal final-state momentum of Beliaev decay $q_\text{min}$ (see Fig. \ref{fig:2nd}b, red dashed line). The momentum distribution can be detected through an angle-resolved measurement of the far-field emission, as sketched in Fig. \ref{fig:setup}a): a photon with (adimensional) in-plane momentum $k$, will fly out of the cavity array at an angle given by $\sin{ \theta_k } = c k/(\omega_L\Delta x)$, with $c$ the speed of light in vacuum \cite{Iacopo_QFL}. 

For the proposed setup, we have that all quantum fluctuations can be detected by restricting the field of view to a cone of aperture $\theta_\text{max} \approx 23^\circ$. Importantly, the dominant signal of the condensate mode at $k=0$ is concentrated about the perpendicular axis and can be filtered out through post-selection.

The momentum-space density of photons escaping from the cavity array is approximately given by $\frac{d\Phi}{dk} =  L\, n_k \gamma / (2\pi)$ where $2\pi/L$ is the momentum-space separation between adjacent modes for an array of $L$ cavities. For an array of $L=128$ cavities, an angular resolution of $\delta k = 0.025\, (2\pi)$  larger than the $k$-space mode separation but well smaller than the width of the Beliaev features, and $n_k \approx 0.1$ around $q_\text{min}$ (see Fig. \ref{fig:2nd}b) we expect a significant photon flux of about $ \Phi = 1.5 \cdot 10^{10}\,$s$^{-1}$. The number of photon clicks per time unit is then given by $ N = \varepsilon_\text{eff} \Phi $ with $\varepsilon_\text{eff}$ some overall efficiency factor incorporating uncontrolled photon losses and detection efficiency. The signal can be integrated in time until a sufficient amount of photons is collected.

Of course, as seen in Fig. \ref{fig:2nd}b, the experimental signal from Beliaev-Landau processes is enhanced with a larger nonlinearity. Although experimentally challenging, a stronger nonlinearity can in principle be achieved by reducing the size of the microcavities or by increasing the excitonic fraction of polaritons \cite{Rodriguez_probing}. Another more speculative possibility is to use the platform of superconducting circuits, where high nonlinearities are naturally achieved \cite{fitzpatrick_PT}.

A crucial point of concern is that the Beliaev-Landau peaks, being rather small in size, can be washed away by a sufficient amount of disorder. In particular, when a small random potential $V_i$ is applied, for instance by variations of the cavity resonance $\omega_c$ from site to site, it will perturb the momentum distribution and imprint additional peaks. We can estimate that the disorder amplitude has to satisfy
$\sqrt{\langle V_i^2 \rangle} \lesssim \hbar\omega_k^\text{peak}\sqrt{\delta n_k^\text{peak}/n_0} \approx 3$ $\mu eV$, with $\omega_k^\text{peak}$ the frequency of the mode at the Beliaev-Landau peak and $\delta n_k^\text{peak}$ the height of the peak (on the order of $2\cdot10^{-3}$, see the red line in the inset of Fig. \ref{fig:2nd}b) ). In Appendix \ref{app:disorder} we provide more details about the derivation of this estimation. In a recent experiment with a setup similar to ours, a standard deviation of about $30\mu$eV for the disorder potential was reported \cite{baboux_unstable}, a factor of about 10 larger than required for our estimations. However, given that the origin of the Beliaev-Landau peaks is different in nature than the disorder background, there are two additional strategies one can employ to isolate them.

First of all, the disorder peaks are different for each realization of a cavity array, while the Beliaev-Landau signal should not depend on this. If one therefore fabricates many copies of the same cavity array on the same sample, averaging over the different copies will cancel out effects from disorder, while the Beliaev-Landau peaks are left in place. If the different copies are positioned adjacent to one another, this amounts to displacing the laser beam from one array to the next.

Secondly, the exact position of the Beliaev-Landau peaks has a well-defined dependence on the mean-field parameters $Un_0$, $\Delta$ and $J$, as plotted in Fig. \ref{fig:bel_decay}c). As a simple example, one could e.g. try to follow the shift of Beliaev-Landau peaks while varying the detuning $\Delta$ by modifying the laser frequency $\omega_L$ according to (\ref{eq:delta}).

Finally, an alternative and conceptually more sophisticated strategy to observe Beliaev-Landau scattering processes is proposed in Fig. \ref{fig:setup}, where we present a sketch of a possible optical setup to directly measure the third-order correlator $M_{k,q}$ (see Fig. \ref{fig:3rd}). The measurement would consist of detecting subtle correlations between the relative phase of the emissions at $k$ and $q$ and a homodyne measurement on the $k-q$ emission mixed with the coherent pump. The detection of simultaneous clicks in detector 1 and the difference signal of detectors 2 and 3 provides a measurement of the quantity
\begin{equation}
\label{eq:detection}
\begin{split}
 \Big \langle \big(\hat{\phi}_q^\dagger + e^{-i\theta}\hat{\phi}_{k}^\dagger \big)\big(\hat{\phi}_q + e^{i\theta}\hat{\phi}_{k} \big)
\big( e^{i\chi} \Omega \hat{\phi}_{k-q}^\dagger +  e^{-i\chi}\Omega^*\hat{\phi}_{k-q} \big) \Big \rangle \\
 = 2 \Re\Big\{ \Omega e^{i(\theta+\chi)}\langle \hat{\phi}_{k-q}^\dagger \hat{\phi}_q^\dagger  \hat{\phi}_k \rangle \Big\} = 2\Re\Big\{ \Omega e^{i(\theta+\chi)}M_{k,q} \Big\},
 \end{split}
\end{equation}
where the second step is obtained after omitting all correlations that are not momentum-conserving, since they must be zero in a spatially uniform sample. The phases $\theta$ and $\phi$ are introduced by the two phase shifters in the setup and allow for the measurement of different quadratures of $M_{k,q}$. Any deviation from zero of the quantity (\ref{eq:detection}) at non-vanishing angles $k,q,k-q \neq 0$ would be a manifest indication of the non-Gaussian nature of the cavity output field and would provide an indication of quasiresonant Beliaev-Landau scattering. 

As high-order interference experiments of this kind go beyond standard quantum optical set-ups, a quantitative study of the expected signal and noise for a realistic experimental setup lies outside the scope of this work and will be the subject of a future study.

\section{Conclusions}
\label{sec:conclusion}

\begin{figure}
\centering
\includegraphics[width = \columnwidth]{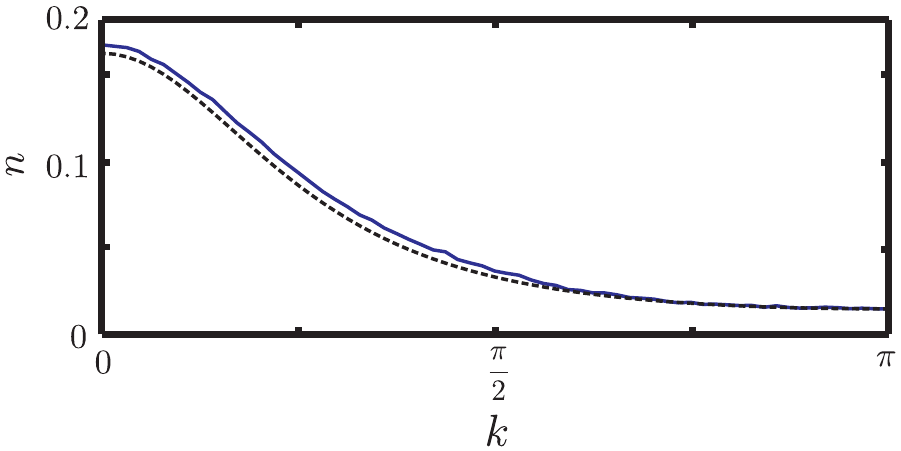} 
\caption{The momentum distribution (full blue line) obtained from a TWA simulation with $(J,\Delta, Un_0) = (10\gamma,-10\gamma, 10\gamma)$ and an interaction constant $U=0.1\gamma$. The result lies very close to the Bogoliubov prediction (dotted black line) when the spectrum does not allow for resonant Beliaev-Landau channels.}
\label{fig:TWA_app}
\end{figure}

In this work we have theoretically studied the effect of Beliaev-Landau processes in a coherently driven fluid of light in a one-dimensional array of weakly nonlinear optical or microwave cavities. In contrast to the equilibrium case, where one typically looks at the decay of additional excitations externally generated in the fluid, here characteristic and experimentally accessible signatures of the Beliaev-Landau processes are identified in observable properties of the nonequilibrium steady-state. 

Remarkably, the momentum distribution (visible in the angular distribution of the far-field emission pattern) shows a characteristic series of peaks and dips, which we attribute to the absence of detailed balance in an out-of-equilibrium setup. Also the higher-order correlators of the field (visible as non-Gaussian features in the photon statistics of the emitted light) exhibit nontrivial features stemming from quasiparticle interactions. Supported by our estimations, we expect that the predicted signal is within the reach of state-of-the-art experimental setups of coupled-cavity arrays with semiconductor microcavities or superconducting circuits.

From the theoretical point of view, our results pinpoint unexpected limitations to the use of the truncated Wigner method to describe scattering processes between quasi-particles. Given the importance of the TWA method as a tool for numerical studies of quantum fluctuation phenomena, future work will address improved schemes to overcome these difficulties. 

The calculations are performed by truncating the hierarchy of correlations of the driven-dissipative Bose-Hubbard model to one order beyond Bogoliubov, i.e. by including the third-order correlation functions and employing a consistent truncation and factorization scheme for the higher-order correlation functions. Future work will extend this technique to spatially inhomogeneous configurations presently of great interest in the context of analog models of gravity.
 
 \acknowledgments
MVR gratefully acknowledges support in the form of a Ph. D. fellowship of the Research Foundation - Flanders (FWO) and hospitality at the BEC Center in Trento. WC and MW acknowledge financial support from the FWO-Odysseus program.
IC was funded by the EU-FET Proactive grant AQuS, Project No. 640800, and by Provincia Autonoma di Trento, partially through the project ``On silicon chip quantum optics for quantum computing and secure communications (SiQuro)''.

\appendix
\section{TWA without Beliaev-Landau channels}
\label{app:TWA}

In the main text we have pointed out how Beliaev-Landau scattering processes lie at the basis of the failure of the truncated Wigner method. To motivate this statement better, we present a TWA simulation of a system which does not contain resonant Beliaev-Landau channels that fulfil condition (\ref{eq:E_cons}). As illustrated in Fig. \ref{fig:bel_decay}, the Bogoliubov spectrum (\ref{eq:bog_transform}) determines the contour of resonant third-order scattering processes. By modifying the mean-field parameters $Un_0$, $J$ and/or $\delta$ we can enter into a regime where no resonant third-order scattering exists. Here we simply choose to replace the value $J=30\gamma$ that was used throughout the main text with $J=10\gamma$, so to have $\Delta < \Delta_0$ ( see Fig. \ref{fig:bel_decay}c)).

As can be seen in Fig. \ref{fig:TWA_app}, the absence of on-shell Beliaev-Landau channels leads to a much better agreement with the prediction 
of the Bogoliubov approximation and does not suffer from unphysical negative densities.  The small deviation from the Bogoliubov distribution can probably be attributed to the nonresonant scattering of the $1/2$ vacuum noise.

\section{Derivation of the correlation hierarchy}
\label{app:HOC}
With ansatz (\ref{eq:ansatz}) from the main text, we find the equation of motion for the quantum fluctuations $\hat{\phi}_k$ (\ref{eq:q_fluct}). By repeatedly applying the product rule, one can obtain the equations of motion for the correlation functions of the quantum fluctuations. An alternative, completely equivalent approach would be to evaluate $\partial_t \langle \hat{O}\rangle = i\big\langle\big[\hat{H},\hat{O}\big]\big\rangle + \text{tr}\big\{\hat{O}\mathcal{D}[\hat{\rho}]\big\}$. Due to spatial homogeneity, only momentum-conserving operator products are included in this construction.

\begin{widetext}
For a general correlation function $C=\left\langle \prod_{k}\hat{\phi}_{k}^{\dagger a_{k}}\hat{\phi}_{k}^{b_{k}}\right\rangle $ of order $N$, with $N = \sum_k \big(a_k + b_k\big)$,  one can derive the following recurrence relation

\begin{align*}
i\frac{\partial C}{\partial t} & =\sum_{q}\left[\left(-\left(a_{q}-b_{q}\right)\left(\epsilon_{q}+2U\left|\psi_{0}\right|^{2}\right)-i\left(a_{q}+b_{q}\right)\frac{\gamma}{2}\right)C\right]\\
& +U\psi_{0}^{2}\sum_{q}\left[\left(2b_{q}C\left[\begin{array}{c}
a_{-q}+\\
b_{q}-
\end{array}\right]+b_{-q}\left(b_{q}-\delta_{q,-q}\right)C\left[\begin{array}{c}
b_{q}-\\
b_{-q}-
\end{array}\right]\right)\right]\\
& -U\psi_{0}^{*2}\sum_{q}\left(2a_{q}C\left[\begin{array}{c}
a_{q}-\\
b_{-q}+
\end{array}\right]+a_{q}\left(a_{-q}-\delta_{q,-q}\right)C\left[\begin{array}{c}
a_{-q}-\\
a_{q}-
\end{array}\right]\right)\\
& +\frac{U\psi_{0}}{\sqrt{L}}\sum_{k,q}\left[2b_{q}C\left[\begin{array}{c}
b_{q}-\\
a_{k-q}+\\
b_{k}+
\end{array}\right]+b_{k}\left(b_{q}-\delta_{q,k}\right)C\left[\begin{array}{c}
b_{k}-\\
b_{q}-\\
b_{k+q}+
\end{array}\right]-a_{k}C\left[\begin{array}{c}
a_{k-q}+\\
a_{q}+\\
a_{k}-
\end{array}\right]\right]\\
& -\frac{U\psi_{0}^{*}}{\sqrt{L}}\sum_{k,q}\left[2a_{q}C\left[\begin{array}{c}
a_{k+q}+\\
a_{q}-\\
b_{k}-
\end{array}\right]+a_{k}\left(a_{q}-\delta_{q,k}\right)C\left[\begin{array}{c}
a_{k+q}+\\
a_{k}-\\
a_{q}-
\end{array}\right]-b_{k}C\left[\begin{array}{c}
b_{k}-\\
b_{q}+\\
b_{k-q}+
\end{array}\right]\right]\\
& +\frac{U}{L}\sum_{k,k',q}\left(2b_{k}C\left[\begin{array}{c}
b_{k}-\\
a_{k'}+\\
b_{k-q}+\\
b_{k'+q}+
\end{array}\right]+b_{k'}\left(b_{k}-\delta_{k,k'}\right)C\left[\begin{array}{c}
b_{k'}-\\
b_{k}-\\
b_{k-q}+\\
b_{k'+q}+
\end{array}\right]-2a_{k}C\left[\begin{array}{c}
a_{k}-\\
a_{k'-q}+\\
a_{k+q}+\\
b_{k'}+
\end{array}\right]-a_{k'}\left(a_{k}-\delta_{k,k'}\right)C\left[\begin{array}{c}
a_{k'}-\\
a_{k'-q}+\\
a_{k+q}+\\
a_{k}-
\end{array}\right]\right)
\end{align*}
Here we adopted the notation, following Ref. \cite{wim_hierarchy}
\[
C\left[a_{q}\pm\right]=\left\langle \hat{\phi}_{q}^{\dagger a_{q}\pm1}\hat{\phi}_{q}^{b_{q}}\prod_{k\neq q}\hat{\phi}_{k}^{\dagger a_{k}}\hat{\phi}_{k}^{b_{k}}\right\rangle
\]

Up to third order, the explicit evaluation of the expression above yields the following equations of motion for the correlators.
\begin{itemize}
\item \emph{First order: }
A finite value for the zero-momentum component is found
\begin{equation}
\label{eq:phi0}
\partial_t \langle \hat{\phi}_0 \rangle = ( Un_0 - i\gamma/2)\langle\hat{\phi}_0\rangle + U\psi_0^2 \langle \hat{\phi}^\dagger_{0} \rangle +   \frac{2U\psi_0}{\sqrt{L}}\sum_k \langle \hat{\phi}^\dagger_{k} \hat{\phi}_k\rangle  + \frac{U\psi_0^\ast}{\sqrt{L}}\sum_k\langle \hat{\phi}_{k} \hat{\phi}_{-k}\rangle +\frac{U}{L}\sum_{k,q} \langle \hat{\phi}^\dagger_{k+q} \hat{\phi}_q \hat{\phi}_k \rangle
\end{equation}

\item \emph{Second order: }
We find for the density of fluctuations
\begin{eqnarray}
i\partial_t \langle \hat{\phi}_k^\dagger \hat{\phi}_k \rangle &=& -i\gamma n_k + U\psi_0^{ 2} \langle \hat{\phi}_k^\dagger \hat{\phi}_{-k}^\dagger \rangle - U\psi_0^{\ast 2} \langle \hat{\phi}_k \hat{\phi}_{-k} \rangle \\
&& +\frac{2U}{\sqrt{L}}\sum_q{\Big( \psi_0 \langle \hat{\phi}^\dagger_k \hat{\phi}^\dagger_q \hat{\phi}_{k+q} \rangle - \psi_0^\ast\langle \hat{\phi}^\dagger_{k+q}\hat{\phi}_q \hat{\phi}_k \rangle \Big)}+ \frac{U}{\sqrt{L}} \sum_q{\Big( \psi^\ast_0 \langle \hat{\phi}^\dagger_k  \hat{\phi}_q \hat{\phi}_{k-q}\rangle - \psi_0 \langle \hat{\phi}^\dagger_{k-q} \hat{\phi}^\dagger_q \hat{\phi}_k \rangle\Big)}\\\label{eq:n_4th}
&&+\frac{U}{L} \sum_{q,l}\Big( \langle \hat{\phi}_k^\dagger \hat{\phi}^\dagger_q \hat{\phi}_l \hat{\phi}_{k+q-l} \rangle - \langle \hat{\phi}_{k+q-l}^\dagger \hat{\phi}_l^\dagger \hat{\phi}_q \hat{\phi}_k \rangle \Big)
\end{eqnarray}
Likewise, for the anomalous averages
\begin{eqnarray}
i\partial_t \langle \hat{\phi}_k \hat{\phi}_{-k} \rangle &=& \big(2\epsilon_k+2U|\psi_0|^2 -i\gamma\big) \langle \hat{\phi}_k \hat{\phi}_{-k} \rangle + U\psi_0^2\bigg(2 \langle \hat{\phi}_k^\dagger \hat{\phi}_k \rangle  + 1\bigg)\\
&& +\frac{2U\psi_0}{\sqrt{L}}\sum_q{ \Big( \langle \hat{\phi}_k \hat{\phi}^\dagger_q\hat{\phi}_{q-k}\rangle + \langle \hat{\phi}^\dagger_q \hat{\phi}_{q+k}\hat{\phi}_{-k} \rangle \Big) } +\frac{U\psi_0^\ast}{\sqrt{L}}\sum_q{ \Big( \langle \hat{\phi}_k \hat{\phi}_q \hat{\phi}_{-k-q} \rangle + \langle \hat{\phi}_q \hat{\phi}_{k-q}\hat{\phi}_{-k}\rangle \Big)}\\\label{eq:c_4th}
 && + \frac{U}{L} \sum_{q,m}{ \Big( \langle \phi_k \hat{\phi}^\dagger_q \hat{\phi}_l \hat{\phi}_{-k+q-l} \rangle + \langle \hat{\phi}^\dagger_q \hat{\phi}_l \hat{\phi}_{k+q-l} \phi_{-k}\rangle \Big)} 
\end{eqnarray}
\item \emph{Third order: }
For the third-order correlation functions we can derive the equations of motion in the same way
\begin{eqnarray}
\nonumber
i\partial_t \langle \hat{\phi}_{k-q}^\dagger \hat{\phi}_q^\dagger \hat{\phi}_{k} \rangle &=&  \left( - \epsilon_{k-q} - \epsilon_q + \epsilon_k - U|\psi_0|^2-\frac{3i}{2}\gamma \right) \langle \hat{\phi}_{k-q}^\dagger \hat{\phi}_q^\dagger \hat{\phi}_{k} \rangle \\
&& -U\psi_0^{\ast 2} \Big( \langle \hat{\phi}_{q-k} \hat{\phi}^\dagger_q \hat{\phi}_k \rangle + \langle \hat{\phi}^\dagger_{k-q} \hat{\phi}_{-q} \hat{\phi}_k \rangle  \Big)+ U\psi_0^2\langle \hat{\phi}_{k-q}^\dagger \hat{\phi}_q^\dagger \hat{\phi}_{-k}^\dagger\rangle \\\nonumber
&& + \frac{2U}{\sqrt{L}} \sum_m \Big( \psi_0 \langle \hat{\phi}^\dagger_{k-q} \hat{\phi}_{q}^\dagger \hat{\phi}_m^\dagger \hat{\phi}_{m+k} \rangle   -\psi_0^\ast \langle \hat{\phi}^\dagger_{k-q+m}\hat{\phi}_m \hat{\phi}_q^\dagger \hat{\phi}_k \rangle - \psi_0^\ast \langle \hat{\phi}^\dagger_{k-q}\hat{\phi}_{q+m}^\dagger \hat{\phi}_m \hat{\phi}_k \rangle  \Big)\\ \label{eq:M_4th}
&& \frac{U}{\sqrt{L}} \sum_m \Big( \psi_0^\ast \langle \hat{\phi}^\dagger_{k-q} \hat{\phi}_{q}^\dagger \hat{\phi}_m \hat{\phi}_{k-m} \rangle    -\psi_0 \langle \hat{\phi}^\dagger_{k-q-m}\hat{\phi}_m^\dagger \hat{\phi}_q^\dagger \hat{\phi}_k \rangle - \psi_0 \langle \hat{\phi}^\dagger_{k-q}\hat{\phi}_{q-m}^\dagger \hat{\phi}_m^\dagger \hat{\phi}_k \rangle  \Big)\\
\label{eq:M_5th}
&& + \frac{U}{L} \sum \Big(\text{5th-order correlators} \Big)
\end{eqnarray}
and
\begin{eqnarray}
\nonumber
i\partial_t \langle \hat{\phi}_{-k-q} \hat{\phi}_q \hat{\phi}_k \rangle &=& \left( \epsilon_k +\epsilon_q + \epsilon_{k+q} + 3U|\psi_0|^2 -\frac{3i}{2}\gamma\right) \langle \hat{\phi}_{-k-q} \hat{\phi}_q \hat{\phi}_k \rangle \\
&& +U\psi_0^2\Big( \langle \hat{\phi}_{k+q}^\dagger \hat{\phi}_q \hat{\phi}_k \rangle +\langle \hat{\phi}_{-k-q} \hat{\phi}_{-q}^\dagger \hat{\phi}_k \rangle +\langle \hat{\phi}_{-k-q} \hat{\phi}_q \hat{\phi}_{-k}^\dagger \rangle \Big)\\\nonumber
&& \frac{2U\psi_0}{\sqrt{L}} \sum_m \Big( \langle \hat{\phi}_{m}^\dagger \hat{\phi}_{m-k-q} \hat{\phi}_q \hat{\phi}_k \rangle +\langle \hat{\phi}_{-k-q} \hat{\phi}_{m}^\dagger \hat{\phi}_{m+q} \hat{\phi}_k \rangle +\langle \hat{\phi}_{-k-q} \hat{\phi}_q \hat{\phi}_{m}^\dagger \phi_{m+k} \rangle \Big)\\ \label{eq:R_4th}
&& \frac{U\psi_0^\ast}{\sqrt{L}} \sum_m \Big( \langle \hat{\phi}_{m} \hat{\phi}_{-k-q-m} \hat{\phi}_q \hat{\phi}_k \rangle +\langle \hat{\phi}_{-k-q} \hat{\phi}_{m} \hat{\phi}_{q-m} \hat{\phi}_k \rangle +\langle \hat{\phi}_{-k-q} \hat{\phi}_q \hat{\phi}_{m} \phi_{k-m} \rangle \Big) \label{eq:R_5th}\\
&& + \frac{U}{L} \sum \Big(\text{5th-order correlators} \Big)
\end{eqnarray}
\end{itemize} 

Through its equations of motion, a correlator of order $N$ couples to correlators up to order $N+2$. In principle, this hierarchy of equations continues to infinite order if the number of particles is not conserved. Therefore, to obtain a closed set of equations, the hierarchy must be truncated in some way.
\end{widetext}

\section{Different truncation schemes and comparison}
\label{app:comp}

\begin{figure}
\centering
\includegraphics[width = \columnwidth]{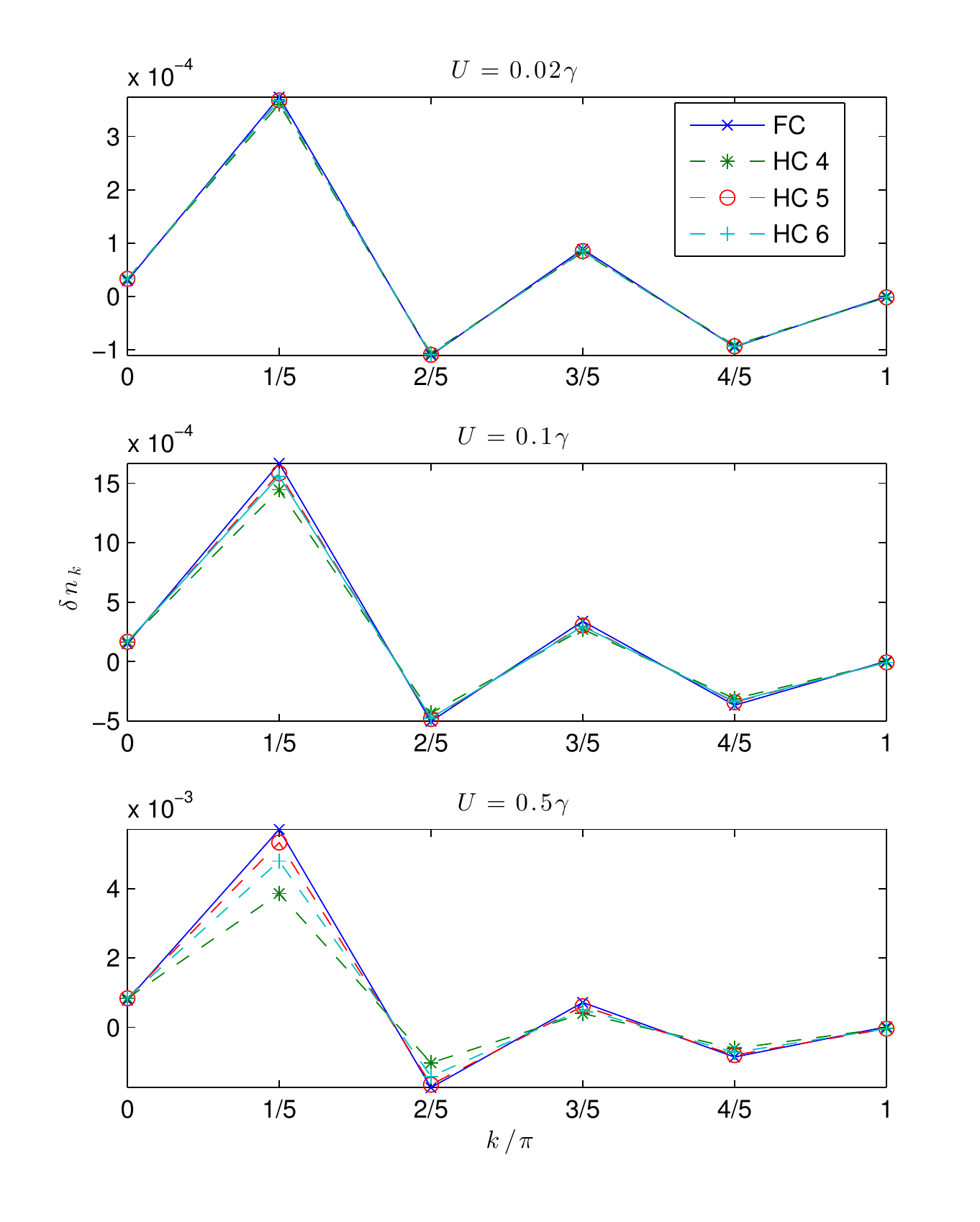} 
\caption{The comparison of different truncation schemes for the quantity $\delta n_k = n_k - n_k^{\text{bog}}$, with $n_k^{\text{bog}}$ the momentum distribution in the Bogoliubov approximation (\ref{eq:momentum_distr}). We show the values of $U$ that have been studied in the main text. The chain consists of only 10 cavities, such that higher-order truncations can be computed and compared. `FC' stands for `factorized cutoff': the method that has been outlined and employed in the main text. `HC' stands for `hard cutoff' and the integer indicates up to which order $N_c$ normal-ordered correlation functions have been included in the hierarchy.}
\label{fig:Kcomp}
\end{figure}

\begin{figure}
\centering
\includegraphics[width = \columnwidth]{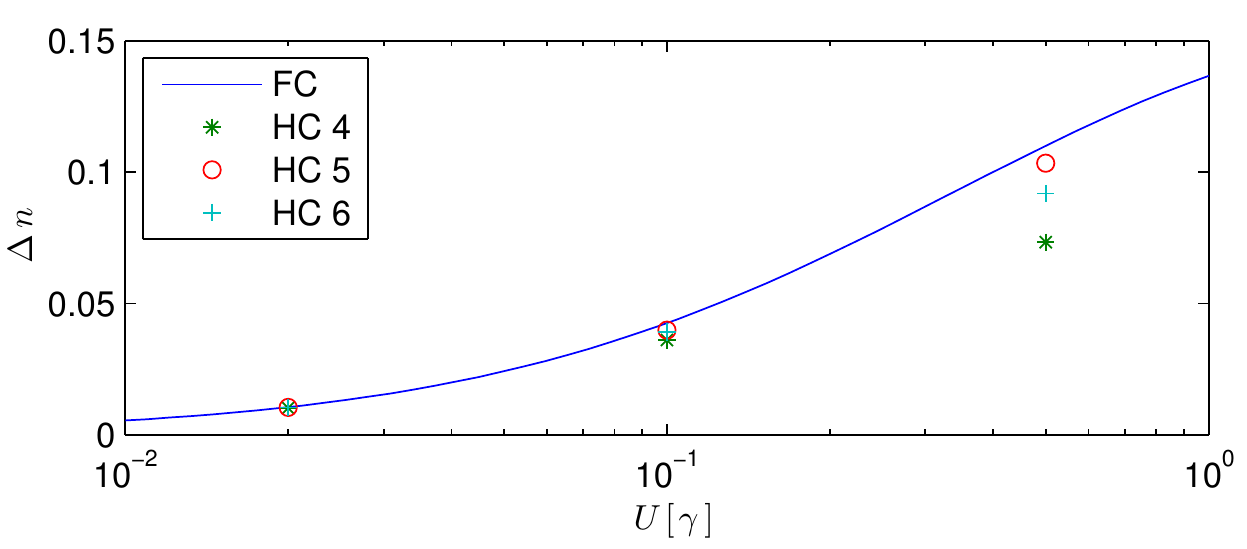} 
\caption{The difference $\Delta n = 1/L \sum_k |n_k - n_k^{\text{bog}}|/n_k^{\text{bog}}$ as a function of the coupling constant $U$ for different truncation schemes for a chain comprising 10 cavities. }
\label{fig:Ucomp}
\end{figure}

We briefly discuss two different truncation schemes and motivate the choice for the one presented in the main text.
\begin{itemize}
\item \emph{The hard cutoff (HC):} The most straightforward approach is to set all correlation functions of order $N$ higher than $N_c$ (i.e. orders $N_c+1$ and $N_c+2$) to zero in the equations of motion. The major benefit of employing this truncation scheme is that, by construction, it produces a linear system of equations, which is relatively easily solved numerically. This allows one to evaluate the result as a function of $N_c$ for small enough system sizes. As discussed in Ref. \cite{wim_hierarchy}, this approach is expected to be efficient when the number of excitations is small. More precisely, by pursuing this truncation scheme, one is implicitly assuming a small occupation of all modes $n_k<1$, otherwise the factorizable part of the correlation functions quickly grows as a function of $N$ and the calculation may not converge at high $N_c$. Nevertheless, even when $n_k < 1$ is satisfied, no a priori assumptions can be made about the connected part of the correlation functions. Therefore one must always verify convergence of the result by increasing $N_c$. Following Ref. \cite{wim_hierarchy}, we refer to this truncation scheme as the `hard cutoff'.

\item \emph{The factorized cutoff (FC):} The method employed in the manuscript differs in a few aspects from the one described above. First of all, we do not use the linearized equation for the first-order correlator $\langle \hat{\phi}_0 \rangle$ (\ref{eq:phi0}), but integrate in time the full Gross-Pitaevskii equation with back-reaction terms, i.e. Eq. (\ref{eq:HE_MF}), along with the different correlation functions. As explained in the main text, this has the advantage of having $\langle \hat{\phi}_0 \rangle=0$ by definition, making correlators up to order $3$ de facto connected, because their factorizable part vanishes. On the other hand, the field $\psi_0$ in equations (\ref{eq:HE_n})-(\ref{eq:HE_R}) is now time dependent, so that the system of equations is no longer linear.

Furthermore, we perform two different approximations to close the system of equations at order $N=3$, by consistently including the back-reaction of the $N=4$ and $N=5$ in the equations for the $N=2$ and $N=3$ correlation functions. While the normal parts of higher order correlators were set to zero in the HC scheme, we start here by including them in factorized form

For a general $N=4$ correlation function, bearing in mind that $\langle \hat{\phi}_0 \rangle=0$, we find that it can be written as,
\begin{equation}
\begin{split}
\langle \phi^\dagger_{m}&\phi^\dagger_{l} \phi_q\phi_k \rangle = \langle \phi^\dagger_{m}\phi^\dagger_{l} \phi_q\phi_k \rangle_c\\
 &+\langle  \phi^\dagger_{l}\phi^\dagger_{-l} \rangle \langle \phi_k\phi_{-k} \rangle \delta_{l,-m}\delta_{k,-q}\\\nonumber
 &+  \langle  \phi^\dagger_{q}\phi_{q} \rangle \langle \phi^\dagger_k\phi_k \rangle\Big( \delta_{m,q}\delta_{l,k} +  \delta_{m,k}\delta_{l,q} \Big) 
\end{split}
\end{equation}
and likewise for other fourth-order correlators. The subscript $c$ (first line) denotes the connected, non-factorizable part of a correlator and is neglected in our truncation scheme. In the equations of motion for the second-order correlation function the factorization of the $N=4$ correlators from (\ref{eq:n_4th}) and (\ref{eq:c_4th}) into products of $N=2$ correlators yields the Hartree-Fock-Bogoliubov-like terms in (\ref{eq:HE_n})-(\ref{eq:HE_c}). In turn, for the third-order, the $N=4$ correlators, given in (\ref{eq:M_4th})- (\ref{eq:R_4th}), produce the drive terms $F^{(M,R)}_{k,q}$ from (\ref{eq:F_M})-(\ref{eq:F_R}). Note that in the latter we have omitted factorizations of the form $\sim \frac{U\psi_0}{\sqrt{L}}c_q \sum_l n_l \delta_{k,0}$, and similar terms with $\sim \delta_{k,q}$ and $\sim \delta_{q,0}$, which drive the diagonal terms of $M(R)_{k,q}$. They are not related to Beliaev-Landau scatterings and we have checked that they merely give a negligible extra shift to $\psi_0$ and slightly renormalize the value of $n_k$ and $c_k$ in $k=0$, while leaving points at $k\neq0$ essentially unaffected.

Also the fifth-order correlator, entering in the equations of motion for the third-order correlator, can be approximated by its factorizable form, which produces a total of ten different products of 2nd and 3rd order correlators.
Two different groups of terms arise with this procedure. For instance, the first of the three terms entering on line (\ref{eq:M_5th}) is
\begin{equation}
\begin{split}
\sum_{l,m}&\langle \hat{\phi}^\dagger_{k-q} \hat{\phi}^\dagger_q \hat{\phi}^\dagger_l  \hat{\phi}_m \hat{\phi}_{k+l-m} \rangle \approx \\
& \langle \hat{\phi}^\dagger_{k-q} \hat{\phi}_{k-q} \rangle  \sum_l \langle \hat{\phi}^\dagger_q \hat{\phi}^\dagger_l \hat{\phi}_{l-q} \rangle + \dots\\
& +  \langle \hat{\phi}^\dagger_{k-q} \hat{\phi}^\dagger_q \hat{\phi}_k \rangle \sum_l \langle \hat{\phi}^\dagger_l \hat{\phi}_l \rangle + \dots
\end{split}
\end{equation}
The dots indicate more terms of  the same kind, with summations over a $N=3$  (first line) or a $N=2$ correlator (second line).
Hence we conclude that, after gathering all those terms, they can be captured by making changes of the kind
\begin{eqnarray}
|\psi_0|^2 &\rightarrow& |\psi_0|^2 + \frac{1}{L}\sum_l \langle \hat{\phi}^\dagger_l \hat{\phi}_l \rangle \\
\psi_0 &\rightarrow& \psi_0 + \frac{1}{L} \sum_l \langle \hat{\phi}^\dagger_q \hat{\phi}^\dagger_l \hat{\phi}_{l-q} \rangle + \dots
\end{eqnarray}
The second approximation consists of neglecting the corrections coming from the factorized fifth-order correlator, which is justified by assuming that the condensate density is much larger than the density of fluctuations. We have evaluated all these fifth-order contributions and verified that their influence on the second-order and third-order correlation function is negligible for the parameters that are used.
\end{itemize}

To check the consistency of the method we used in the main text, we compare it in Fig. \ref{fig:Kcomp} and Fig. \ref{fig:Ucomp} with the HC truncation scheme at higher orders for a chain of only 10 sites. This allows us to obtain results within a reasonable computation time for truncation orders up to $N_c=6$ in the HC scheme. The factorized method from the main text appears to agree very well with the HC scheme for $N_c=5$, even though we only included connected correlators up to order 3. From this we conclude that, at least in the parameter regime that we have considered, it is a good approximation to neglect both the connected $N=4$ and the full $N=5$ correlator. On the other hand, we see a large deviation from the HC with $N_c=4$, even though both methods include correlators up to the same order $N=4$. This can be attributed to the inaccuracy with which the fourth-order is evaluated in the HC scheme, i.e by bluntly neglecting  all higher-orders. Therefore the inclusion of the fourth order in factorized form directly, as was employed in our FC scheme, turns out to be a much better approximation than obtaining it through a HC scheme with $N_c=4$.

From Fig. \ref{fig:Kcomp} and Fig. \ref{fig:Ucomp} we deduce that results obtained with different truncation schemes start to deviate from each other at $U/\gamma=0.5$. This limits the range of parameters in which our approach provides a quantitatively accurate description. While for stronger $U/\gamma$, the error due to truncation of higher orders terms becomes increasingly important, still at $U/\gamma=0.5$ we see that all truncation schemes reproduce at least qualitatively the same result; in particular, the $N_c=3$ factorized cutoff still reasonably well agrees with the $N_c=5$ HC scheme, and even with $N_c=6$. We have therefore chosen to use the $N_c=3$ factorized cut-off approximation for the study of Beliaev-Landau processes in larger systems, even though there is a small, but non-negligible quantitative deviation from higher-order HC schemes. 

Keeping in mind that higher-order HC schemes are numerically cumbersome and can not be applied to large systems, our main motivation for sticking to the $N_c=3$ factorized cut-off approximation is that, to our knowledge, no efficient methods exist to simulate driven-dissipative quantum dynamics in a system with intermediate interactions ($U=0.5\gamma$) and a large particle number (128 cavities, each containing $20$ photons on average). A possible alternative approach would be to develop a variational method with matrix product operators in the spirit of \cite{mascarenhas_matrix,cui_variational}, in which the matrix product state of the quantum excitations is determined self-consistently by coupling it back to the coherent condensate. This will 
hopefully be the subject of future work.

We therefore conclude that, even though we constructed a hierarchy in terms of correlation functions up to order 3 only, the accuracy is comparable to (or even better as) higher-order methods in the HC scheme. Obviously, the reduced numerical complexity in the developed truncation scheme realizes a significant computational speedup as compared to these higher-order methods, thus allowing us to tackle much larger systems and address the physically most relevant questions.

\section{The influence of disorder}
\label{app:disorder}
Given the Fourier transform $V_k$ of a random potential, $V_j = \frac{1}{\sqrt{L}} \sum_k V_k e^{ik j}$ that is applied to the cavity array. We find that the mean field follows the equation of motion
\begin{eqnarray}
\label{eq:disorder_MF}
\nonumber
i \dot{\psi}_j &=& \Big( V_j - \delta - i\frac{\gamma}{2}\Big)\psi_j-J(\psi_{j+1} + \psi_{j-1}) \\&& + U|\psi_j|^2\psi_j + \Omega_j
\end{eqnarray}
 In the linear-response regime, the non-uniform polariton field can be formulated as $\psi_{j} = \psi_0 + \frac{1}{\sqrt{L}}\sum_k \delta\psi_k e^{ikj}$. After substitution in (\ref{eq:disorder_MF}) and collecting terms up to linear order in $\delta \psi_k$ and $V_k$, we derive a linear set of equations for each mode
 \begin{equation}
 \label{eq:linear}
 \mathcal{L}_k \vectr{\delta\psi_k}{\delta\psi_{-k}^\ast} = \vectr{-V_k\psi_0}{V_k\psi^\ast_0}
 \end{equation}
 with the response matrix
 \begin{equation}
  \mathcal{L}_k = \matrx{\epsilon_k + Un_0 -i\frac{\gamma}{2}}{U\psi_0^2}{-U\psi_0^{\ast2}}{-\epsilon_k - Un_0 -i\frac{\gamma}{2}}
 \end{equation}
 and $\epsilon_k$ given in (\ref{eq:epsilon}). Solving (\ref{eq:linear}) yields the response of the density distribution to the disorder potential in the linear regime 
\begin{equation}
\label{eq:n_disorder}
\delta n_k = |\delta\psi_k|^2 = \big| V_k \psi_0 \big|^2 \frac{ \epsilon_k^2 + \gamma^2/4}{ \big(\omega_k^2 + \gamma^2/4\big)^2} 
\end{equation} 
with $\omega_k$ given in (\ref{eq:bog_transform}).
Since all energy scales are larger than $\gamma$ and $\omega_k \approx \epsilon_k$ for the purposes of this qualitative analysis, we can further approximate $\delta n_k \sim n_0 \big( V_k /\omega_k \big)^2$. For white uncorrelated noise it therefore follows that roughly $\sqrt{\langle V_j^2 \rangle} \lesssim \omega_k^\text{peak} \sqrt{ \delta n_k^\text{peak}/n_0}$ if we want the peaks of disorder to be smaller than the peaks of Beliaev-Landau scattering.

\bibliography{bibliography}{}
\end{document}